\documentclass[useAMS,usenatbib]{mn2e}

\usepackage{Times}

\usepackage{psfrag}
\usepackage{pspicture}
\usepackage{graphicx}

\title[Bright Ly$\alpha$ emitters at z$\sim$9]{Bright Ly$\alpha$ Emitters at z$\sim$9: constraints on the luminosity function from HiZELS \thanks{Based on observations obtained with the Wide Field CAMera (WFCAM) and the Cooled Grating Spectrometer (CGS4) on the United Kingdom Infrared Telescope (UKIRT) as part of the Hi-z Emission Line Survey (HiZELS)} }
\author[D. Sobral et al.]{D. Sobral$^{1}$\thanks{E-mail: drss@roe.ac.uk}, P. N. Best$^{1}$, J. E. Geach$^{2}$, Ian Smail$^{2}$, J. Kurk$^{3}$, M. Cirasuolo$^{1,5}$, M. Casali$^{4}$,
\newauthor R. J. Ivison$^{1,5}$, K. Coppin$^{2}$ \& G. B. Dalton$^{6,7}$ \\
$^{1}$SUPA, Institute for Astronomy, Royal Observatory of Edinburgh, Blackford Hill, Edinburgh, EH9 3HJ, UK\\
$^{2}$Institute for Computational Cosmology, Durham University, South Road, Durham, DH1 3LE, UK\\
$^{3}$Max-Planck-Institut f{\"u}r Astronomie, K{\"o}nigstuhl, 17 D-69117, Heidelberg, Germany\\
$^{4}$European Southern Observatory, Karl-Schwarzschild-Strasse 2, D-85738 Garching, Germany\\
$^{5}$Astronomy Technology Centre, Royal Observatory of Edinburgh, Blackford Hill, Edinburgh, EH9 3HJ, UK\\
$^{6}$Astrophysics, Department of Physics, Keble Road, Oxford, OX1 3RH, UK\\
$^{7}$Space Science and Technology, Rutherford Appleton Laboratory, HSIC, Didcot, OX11 0QX, UK}
\begin{document}

\date{Accepted 2009 June 29. Received 2009 June 29; in original form 2009 June 18}

\pagerange{\pageref{firstpage}--\pageref{lastpage}} \pubyear{2009}

\maketitle

\label{firstpage}




\begin{abstract}
New results are presented, as part of the Hi-$z$ Emission Line Survey (HiZELS), from the largest area survey to date (1.4 deg$^2$) for Ly$\alpha$ emitters (LAEs) at $z\sim9$. The survey, which is primarily targeting H$\alpha$ emitters at $z<3$, uses the Wide Field CAMera on the United Kingdom Infrared Telescope and a custom narrow-band filter in the $J$ band and reaches a Ly$\alpha$ luminosity limit of $\sim10^{43.8}$\,erg\,s$^{-1}$ over a co-moving volume of $1.12\times10^6$\,Mpc$^3$ at $z=8.96\pm0.06$. Only 2 candidates were found out of 1517 line emitters and those were rejected as LAEs after follow-up observations. The limit on the space density of bright Ly$\alpha$ emitters is improved by 3 orders of magnitude, consistent with suppression of the bright end of the Ly$\alpha$ luminosity function beyond $z\sim6$. Combined with upper limits from smaller but deeper surveys, this rules out some of the most extreme models for high-redshift Ly$\alpha$ emitters. The potential contamination of future narrow-band Ly$\alpha$ surveys at $z>7$ by Galactic brown dwarf stars is also examined, leading to the conclusion that such contamination may well be significant for searches at $7.7<z<8.0$, $9.1<z<9.5$ and $11.7 < z < 12.2$.

\end{abstract}


\section{Introduction}

One of the most important questions in astronomy is ``when did the first stars and galaxies form?''. Observations of the most distant galaxies offer one of the greatest possible constraints on structure formation, allowing models of early galaxy formation and evolution to be tested, refined or refuted.  Over the last decade, considerable manpower and
telescope time has been dedicated towards this goal: galaxies have now been identified out to redshift $z\sim7$, just 750 Myr after the Big Bang \citep{Iye}, and recently a Gamma Ray Burst (GRB) has been detected even further away, at $z\approx 8.3$ \citep[][]{Tanvir}; the sample of very-high redshift galaxies is growing rapidly. Making the additional step out to redshifts of $z \sim 9$ is of the upmost importance because it not only offers much tighter constraints on the first star formation or AGN activity of the Universe, but also allows the re-ionisation epoch of the Universe to be studied. As the mean fraction of neutral hydrogen in the intergalactic medium increases, the Ly$\alpha$ emission from these star forming galaxies will be strongly attenuated, with dramatic consequences for the shape of the Ly$\alpha$ luminosity function, although the precise details may depend upon the level of local ionization of the intergalactic medium by the star forming galaxies \citep[e.g.][]{Haiman}. Little evolution is seen in the Ly$\alpha$ luminosity function between $z\sim3$ and $z\sim6$, suggesting that the Universe was effectively fully ionized by $z=6$ \citep[e.g.][]{malhotra04,Ouchi08}, although some hints of evolution have been found at the bright end of the luminosity function beyond $z\sim6$ \citep[e.g.][]{Kashikawa06,Ota08}. Still, the re-ionisation epoch is widely believed to occur around $z\sim9$, with this being supported by several models and observations; the latest results from the Cosmic Microwave Background \citep[e.g.][]{CMB,Komatsu} show that the bulk of the re-ionisation occurred at $z=10.9\pm1.4$ ($1\sigma$).

Presently, there are three relatively effective methods for searching for very distant galaxies: the broad-band drop-out technique, ``blind" spectroscopic searches and narrow-band imaging surveys. The widely-used drop-out technique \citep[pioneered at $z\sim3$ by][]{steidel96} requires very deep broad-band imaging, and can potentially identify $z>7$ galaxies as z-band drop-outs \citep[e.g.][]{bouwens08,richard08}. This method is efficient for identifying candidates, but requires detailed spectroscopic follow-up to confirm the candidates, especially to rule out contributions from other populations with large z$-J$ breaks, such as dusty or evolved $z\sim2$ galaxies and ultra-cool galactic stars \citep[e.g.][]{mclure06}. Thus, while \cite{richard08} identified two $z\sim9-10$ candidates by taking advantage of the lensing magnification of a high mass cluster, their spectroscopic follow-up was inconclusive with no emission lines detected. ``Blind" spectroscopic surveys can potentially provide spectra directly. They are always limited to very small areas, although \cite{stark07} targetted the critical lensing lines of clusters and were able to identify 6 potential $z\sim9$ objects. Finally, the narrow-band imaging technique has the advantage of potentially probing very large volumes, but can only detect sources with strong emission lines, whilst it still depends on the Lyman-break technique to isolate very high-redshift emitters.

Narrow-band Ly$\alpha$ searches at slightly lower redshifts have been extremely successful in detecting and confirming emitters \citep[e.g.][]{hu99}, including the detection of the most distant (spectroscopically confirmed) Ly$\alpha$ emitter to date at $z=6.96$ \citep{Iye}. There have been attempts to detect Ly$\alpha$ at $z>7$, and particularly at $z\sim9$ \citep[e.g.][]{willis05,cuby07,willis08}, some taking advantage of cluster lensing magnifications; all such studies have been unsuccessful to date, but have only surveyed very small areas (a few tens of square arcmins at most). With the advent of wide-field near-IR detectors, however, it is now possible to increase the sky areas studied by over 2 to 3 orders of magnitude and reach the regime where one can realistically expect to detect $z \sim 9$ objects. This is a key aim of, for example, the narrow-band component of the UltraVISTA Survey \citep[c.f.][]{Nilsson07}. It is also an aim of HiZELS, the Hi-Z Emission Line Survey \citep[c.f.][]{Geach,Sobral}, that we are carrying out using the WFCAM instrument on the 3.8-m UK Infrared Telescope (UKIRT). HiZELS is using a set of existing and custom-made narrow-band filters in the $J$, $H$ and $K$ bands to detect emission line galaxies over $\sim$ 5 square degrees of extragalactic sky; the narrow-band $J$ filter (hereafter NB$_{\rm J}$) is sensitive to Ly$\alpha$ at $z=8.96$.

In this work, an H$_0=70$\,km\,s$^{-1}$\,Mpc$^{-1}$, $\Omega_M=0.3$ and $\Omega_{\Lambda}=0.7$ cosmology is used; magnitudes are given in the Vega system.

\section{DATA AND SELECTION}

Deep narrow-band $J$ (NB$_{\rm J}\approx21.6$, 3$\sigma$, ${\rm F_{lim}}=7.6\times10^{-17}$\,erg\,s$^{-1}$\,cm$^{-2}$) imaging was obtained across 1.4 deg$^2$ in the UKIRT Infrared Deep Sky Survey Ultra Deep Survey \citep[UKIDSS UDS;][]{2007MNRAS.379.1599L} and the Cosmological Evolution Survey \citep[COSMOS;][]{2007ApJS..172....1S,2007ApJS..172..196K} fields, both of which have a remarkable set of deep multi-wavelength data available -- this resulted in the selection of 1517 potential line emitters. The NB$_{\rm J}$ filter ($\lambda=1.211\pm0.015\mu$m) is sensitive to Ly$\alpha$ emission at $z=8.96\pm0.06$ (assuming a top-hat filter shape), probing a co-moving volume of 1.12$\times10^6$ Mpc$^3$ -- by far the largest probed by a narrow-band survey at these wavelengths. The reader is referred to \cite{Sobral} -- hereafter S09 -- for details regarding the observations, data reduction and the general selection of narrow-band emitters. Here the focus will be on identifying Ly$\alpha$ emitters candidates within that data-set.

\subsection{Search for Candidates}

For a source to be considered a candidate $z\approx9$ Ly$\alpha$ emitter it is required to: i) be selected as a narrow-band emitter in S09 (this required it to be clearly detected in NB$_{\rm J}$ ($\sigma>3$) with a $J$-NB$_{\rm J}$ colour excess significance of $\Sigma>2.5$ and observed equivalent width EW$ >50$ \AA \ -- see S09 for details in which it is shown that these criteria are very robust); ii) have at least one other detection $>3\sigma$ in the near-infrared; iii) be visually believable in NB$_{J}$ and the other band(s), avoiding noisy areas; and iv) be undetected ($<3\sigma$ and direct visual analysis) in the available visible band imaging ($B$,$V$,$r$,$i$,z) -- \sc{subaru} \rm and ACS/\sc{hst}). \rm

The sample presented in S09 was used to search for potential Ly$\alpha$ emitters at $z\approx9$. However, the investigation was also extended to a slightly larger area in the UKIDSS UDS field to include areas where deep SUBARU and near-infrared imaging data were available -- this corresponds to re-including areas which were conservatively masked for SED fitting purposes in S09, and increases the total area probed to $1.4$ deg$^2$.

\subsection{Candidates, testing and follow-up observations}

No candidates were found in the UKIDSS UDS field, with all emitters that passed tests i) to iii) being clearly detected in z-band imaging. In COSMOS, however, 2 candidates were found that satisfied all criteria. Both sources are absent in all optical bands down to the 3$\sigma$ level (e.g. ${\rm I}=28.1$ mag, ${\rm z}=25.8$ mag). They are both detected in NB$_{\rm J}$ and $J$, with Cand 1 having NB$_{\rm J}=20.8$ mag (5$\sigma$) and $J=21.5$ mag (9$\sigma$) and a drop $z-J>4.2$ mag, while Cand2 presents NB$_{\rm J}=$ 20.8 mag (5$\sigma$) and $J=22.1$ mag (6$\sigma$) and a drop $z-J>3.6$ mag. They are both undetected in all other infra-red bands.

These two sources were then subjected to a series of further tests and follow-up observations. Splitting the data into subsets confirmed the detections across observations conducted on different nights (timescale from one day up to one month), with no evidence for variability or proper motion; they were also clearly excluded as potential cross-talk artifacts. Cand1 was followed-up spectroscopically using the CGS4 instrument on UKIRT in January 2009 -- these data failed to confirm an emission line. Both candidates were then re-observed using WFCAM (further $J$ imaging in February 2009), resulting in the non-detection of both candidates. Also, $J$ imaging from the COSMOS public archive (which has become publicly available very recently) fails to detect the candidates. It is therefore clear that these sources are not Ly$\alpha$ emitters at $z\sim9$. Further investigation shows that the sources are likely to be artifacts; they are located at almost the same pixel position on WFCAM4 in 2 different paw-prints (later investigation showed that these were also found at similar pixel positions in the remaining 2 paw-prints, although at a lower level, and thus were not selected). They seem to be caused by an unfortunate coincidence of a set of slightly hot pixels (not sufficient to be flagged as bad pixels) which, combined with the ditter pattern, produced a few $\sigma$ excess at one location on the combined image. No other artifacts like these were found either in UDS (which used a different dither pattern) or in other cameras for COSMOS data.

\section{RESULTS \& DISCUSSION}

%
%
\begin{figure}
\centering
\includegraphics[width=8.2cm]{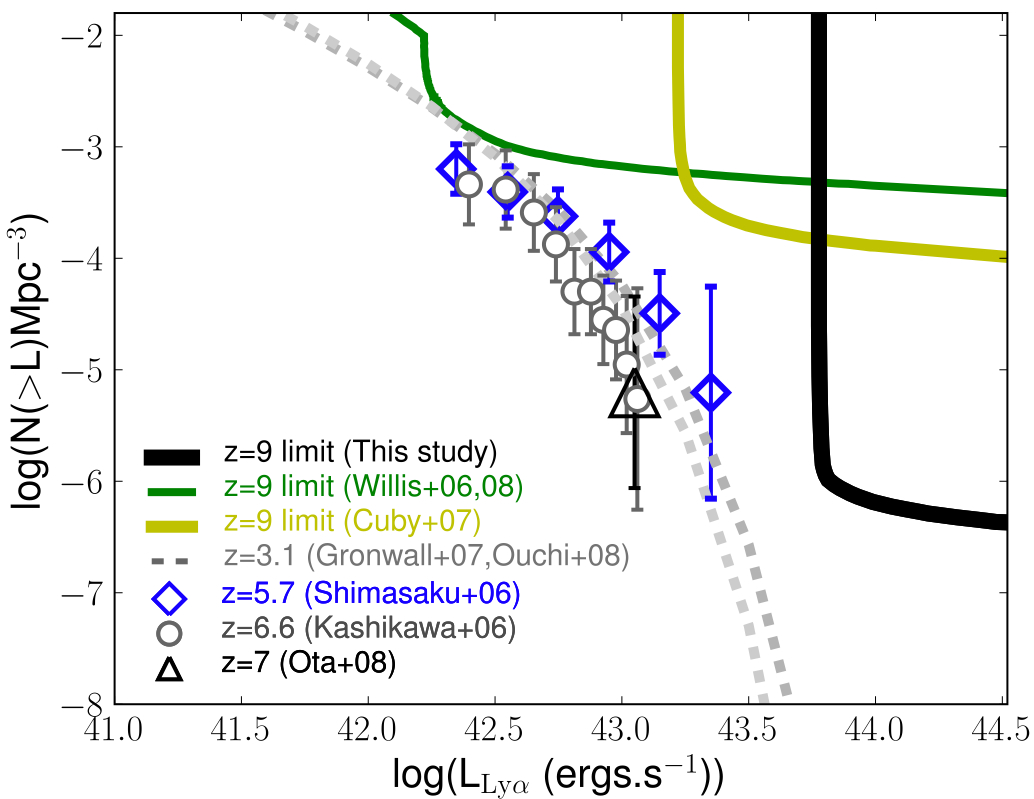}
\caption[LF1]{Comparison between the measured Ly$\alpha$ luminosity function at $z\sim3$ \citep[dotted lines;][]{Gronwall07,Ouchi08} with data from $z\sim6-7$ \citep{Kashikawa06,Shimasaku06,Ota08}. Other typically smaller $z\sim6$ surveys give consistent results within the error bars. No evidence of significant evolution is found, especially when accounting for cosmic variance. Limits for the $z\sim9$ LF from \cite{willis05}, \cite{cuby07} and \cite{willis08} are also presented, together with the one presented in this Letter, which is inconsistent with a strong evolution in L$^*$ up to $z\sim9$.  \label{fig1}}
\end{figure}

\subsection{Ly$\alpha$ luminosity function at $z\sim9$}

After conducting the widest survey of bright (${\rm L}>7.6\times10^{43}$\,erg\,s$^{-1}$) Ly$\alpha$ emitters at $z\sim9$, probing a co-moving volume of 1.12$\times10^6$ Mpc$^3$ (1.4 deg$^2$), only 2 sources passed the selection criteria and even those were ruled out after follow-up observations. This result allows the tightest constraint on the bright end of the $z\sim9$ Ly$\alpha$ luminosity function, as previous surveys \citep{willis05,cuby07,willis08} have only covered very small areas (a factor $\sim$1000 smaller). However, those surveys have gone significantly deeper (up to a factor of $\sim$100). Thus, by combining all the results from the literature, the luminosity function of LAEs at $z\sim9$ can be constrained across a wide range of luminosities: 10$^{42}<L<10^{45}$\,erg\,s$^{-1}$. Figure \ref{fig1} presents the constraints from \cite{willis05}, \cite{cuby07}, \cite{willis08} and from this work, indicating the inverse of the volume selection function for each survey. These are compared to the measured Ly$\alpha$ luminosity functions from $z\sim3$ to $z\sim7$ from recent studies.
Although the samples of Ly$\alpha$ emitters may suffer from significant biases due to the selection, cosmic variance and possible contamination, Figure \ref{fig1} reveals that there is little evolution in the bright end of the luminosity function between $z\sim3$ and $z\sim5.7$. However, those bright emitters seem to become much rarer at $z=6.5$ \citep{Kashikawa06}, indicating that $L^*$ is not increasing from $z\sim6$ onwards. The results presented in this Letter are also consistent with no evolution in $L^*$ ($\Delta$log(L$^*$)$<$0.5) from $z=5.7$ to $z\sim9$.

%
%
\begin{figure}
\centering
\includegraphics[width=8.2cm,height=6.3cm]{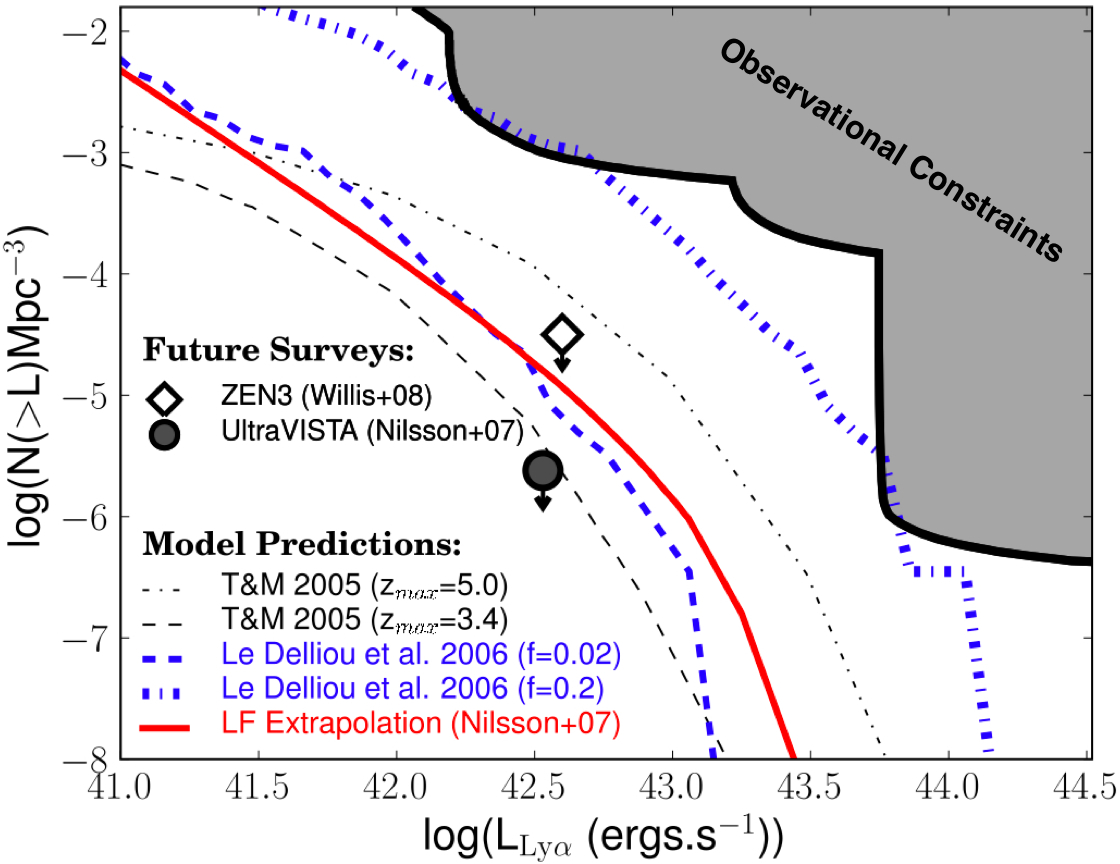}
\caption[LF2]{The observational limits on the $z\sim9$ Ly$\alpha$ luminosity function compared to different model predictions and proposed future surveys. The most recent models agree well with the data limits, and only the most extreme ones can be ruled out. Also, according to these models, the UltraVISTA survey will have a clear chance of detecting a few of these Ly$\alpha$ emitters, while ZEN3 may get a detection and will at least be able to rule out more models.\label{fig2}}
\end{figure}

\subsection{Comparison with models and future surveys}
\label{compl}

Several authors have tried to predict the Ly$\alpha$ luminosity function at $z\sim9$, either by extrapolating the luminosity function of these emitters from lower redshift, or by using numerical or semi-analytical models \citep{Thommes05,LeDelliou06,Nilsson07}. In this Letter, different models are compared with the observational constraints: semi-analytical models, observational extrapolations and phenomenological models. The semi-analytical models discussed here are obtained from {\sc galform} \citep{Baugh05} -- these are based on $\Lambda$CDM, having been successful in reproducing a wide range of galaxy properties at different redshifts, including Ly$\alpha$ emitters up to $z\sim6$ \citep{LeDelliou06}. {\sc galform} computes the build-up of dark matter halos by merging and the assembly of baryonic mass of galaxies and the semi-analytical approach allows the study of properties of the Ly$\alpha$ emission -- the reader is referred to \cite{Baugh05}, \cite{LeDelliou06} and \cite{Orsi} for more details on these. The observational approach, as in \cite{Nilsson07}, extrapolates the Schechter function parameters based on those obtained in the $3.1<z<6.5$ redshift range. In practice, this results in little $L^*$ evolution but a significant negative $\phi^*$ evolution. Finally, the phenomenological approach in \cite{Thommes05} assumes that Ly$\alpha$ emitters at high redshift are spheroids seen during their formation phase. These models are normalised to give the observed mass function of spheroids in the local Universe, and are combined with a phenomenological function that provides the distribution of spheroid formation events in mass and redshift. Each galaxy is assumed to be visible as a Ly$\alpha$ emitter during a starburst phase of fixed duration that occurs at a specific redshift, drawn from a broad distribution. The reader is referred to \cite{Thommes05} for details.

Figure \ref{fig2} presents predictions from {\sc galform} \citep{LeDelliou06}, the observational luminosity function extrapolation from \cite{Nilsson07} and updated phenomenological predictions \citep{Thommes05} assuming peak redshifts of $z_{max}=3.4$ and $z_{max}=5.0$. \cite{LeDelliou06} found that their model required an escape fraction of f$_{\rm esc}=0.02$ to fit the observed Ly$\alpha$ luminosity function at $3<z<6.5$, but they also presented predictions for f$_{\rm esc}=0.2$ to illustrate how the results might change if the escape fraction increased at very high redshift; we therefore show both predictions in Figure \ref{fig2}. While most predictions are consistent with the current limits, {\sc galform} models with high escape fractions are marginally rejected both at faint and bright levels. Earlier phenomenological models \citep[e.g. the $z_{max}=10$ model of][not shown in Figure \ref{fig2}]{Thommes05} are also clearly rejected by our results.

These results also show, observationally, that bright ${\rm L}>10^{43.8}$\,erg\,s$^{-1}$ Ly$\alpha$ emitters are very rare. Although the area coverage is absolutely important, a depth+area combination is likely to be the best approach for gathering the first sample of these very high-redshift galaxies. In fact, that is the strategy of the narrow-band component of the UltraVISTA survey \cite[c.f.][]{Nilsson07}, using the VISTA telescope, which will map 0.9 deg$^2$ of the COSMOS field to a planned 5$\sigma$ flux limit of $4\times10^{-18}$ erg\,s$^{-1}$\,cm$^{-2}$; this corresponds to luminosity limit of $L=10^{42.53}$\,erg\,s$^{-1}$ and a surveyed volume of 5.41$\times10^5$ Mpc$^3$ (see Figure \ref{fig2}) at $z=8.8$. This combination lies below all current predictions for the $z\sim9$ Ly$\alpha$ LF and the survey is expected to detect 2-20 Ly$\alpha$ emitters at $z=8.8\pm0.1$. On the other hand, the ZEN3 survey \citep[c.f.][Hibon et al. 2009, submitted to A\&A]{willis08} will also try to get a compromise between area and depth by taking advantage of the Canada-France-Hawaii Telescope (CFHT) and their near-infrared large area camera (WIRCam); that survey will not go as wide or as deep, and whilst it will provide significantly better constraints, it is not clear (as can be seen in Figure \ref{fig2}) whether or not it will be successful in detecting any Ly$\alpha$ emitter.

\subsection{High redshift Ly$\alpha$ searches and cool galactic stars}

It has become widely realised in recent years that broad-band searches for $z>6$ galaxies using the Lyman-break technique may suffer from significant contamination by cool Galactic L, T, and possibly Y-dwarf stars \citep[e.g.][]{2006MNRAS.372..357M}. These low-mass brown dwarfs display extremely red $z-J$ colours reaching as high as $z-J \approx 4$ \citep[e.g.][]{2008MNRAS.391..320B}, coupled with relatively flat $J-K$ colours. Such colours can mimic very closely those expected of a $z>6$ star forming galaxy with a strong Lyman-break.

\begin{figure}
\centering
\includegraphics[width=8.2cm]{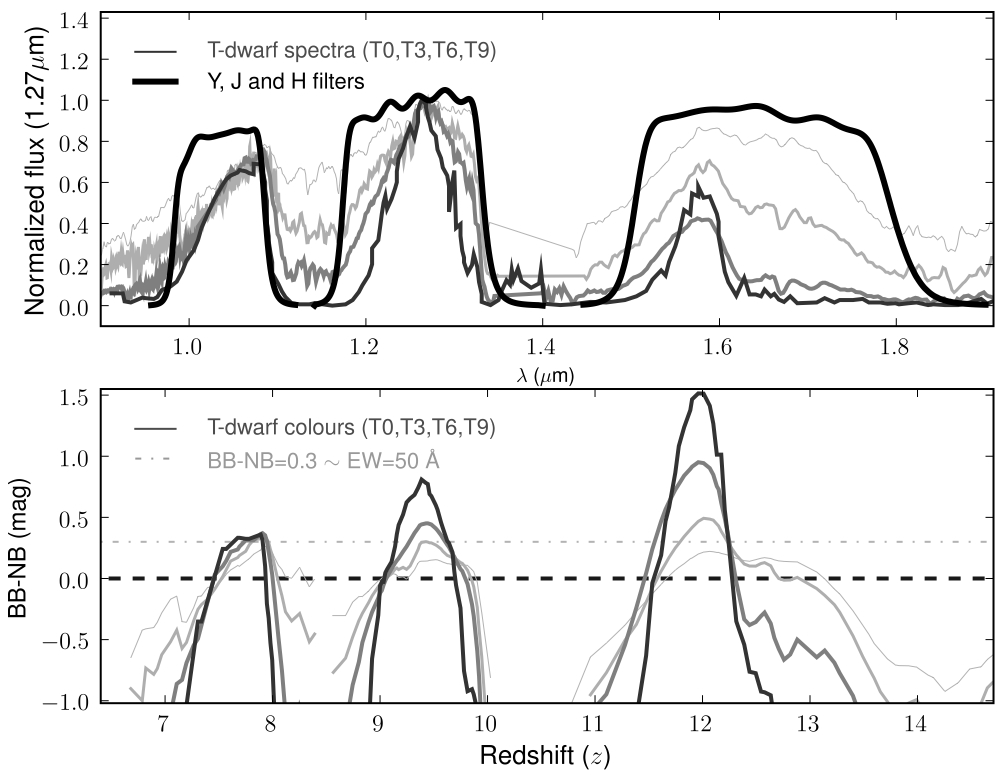}
\caption[Tdwarfs]{{\it Top panel:} the near-infrared spectra of T0, T3, T6 and T9 dwarf stars \citep[T0 -- lighter, T9 -- darker, from][]{2008MNRAS.391..320B}) compared to near-IR broad band filter profiles. {\it Lower panel:} the consequences for measured broad-band minus narrow-band (BB-NB) colours, clearly demonstrating the redshifts/wavelengths at which searches for Ly$\alpha$ emitters can be significantly contaminated by these very cool stars. For $7.7<z<8.0$ and $9.1<z<9.5$ searches, these stars can easily mimic Ly$\alpha$ emitters, with strong $Y$-z or $J$-z breaks and significant positive BB-NB colours. Searches at higher redshift $11.6 < z < 12.2$ in the $H$ band can detect T9s with BB-NB$\sim$1.5, although the lack of strong H-J or H-Y breaks will make it easier to distinguish T-dwarfs from Ly$\alpha$ emitters. \label{fig3}}
\end{figure}

It may be thought that narrow-band Ly$\alpha$ searches are immune to this contamination, since the initial emission-line galaxy selection relies on an excess flux observed in a narrow-band filter relative (usually) to a broad-band filter; only after that is the Lyman-break technique used to pick out the high-redshift Ly$\alpha$ candidates from amongst the emission-line objects. However, the near-infrared continuum spectra of low mass brown dwarfs show considerable structure due to broad molecular absorption features \citep[especially methane and ammonia; e.g.][]{2007ApJ...667..537L}, as shown in the top panel of Figure \ref{fig3}. The lower panel of Figure \ref{fig3} shows very clearly that T-dwarfs can easily produce a positive broad-band minus narrow-band (BB-NB) colour if the narrow-band filter is located within one of the spectral peaks (note that this is much less of an issue for surveys which difference two closely-located narrow-band filters). Ly$\alpha$ narrow-band surveys in the redshift ranges $7.7<z<8.0$, $9.1<z<9.5$ and $11.7 < z < 12.2$ may therefore be prone to contamination by cool Galactic stars -- this includes the $z=7.7$ and $z=9.4$ atmospheric windows for narrow-band searches of Ly$\alpha$ emitters. Narrow-band surveys at redshifts $z<7.7$, or between $8.0<z<9.1$ -- which includes both HiZELS ($z=8.96$) and the narrow-band component of the UltraVISTA Survey ($z=8.8$; e.g. Nilsson et al 2007) -- will be free of such contamination. Indeed, such surveys could potentially select very cool T-dwarf stars via a narrow-band {\it deficit} due to the strong methane absorption feature at these wavelengths.

With this result in mind, a T-dwarf search was conducted among narrow-band {\it deficit} sources in S09. These {\it deficit} sources were selected using equivalent criteria as for emitters (with a change in sign). None of the $deficit$ sources has z(AB)$-J>3$, as expected for T-dwarfs (e.g. Leggett et al. 2007, Burningham et al. 2008), and even a selection imposing z(AB)$-J>2$ results in a sample of only 9 galaxies which are all very well SED-fitted as galaxies with $z_{\rm photo}\sim1.4-1.5$. These sources also present slightly higher $J$ and $H$ fluxes when compared to the best SED fit, but this can be explained by the H$\beta$ and [O{\sc iii}] contributing to the $J$ band and H$\alpha$ to the $H$ band, which also explains the NB$_J$ deficit. No T-dwarf candidate was found in our survey.

%
%

\section{Summary}

\begin{itemize}

\item Deep narrow-band imaging in the $J$ band ($\lambda=1.211\pm0.015\mu$m) has been used to search for bright Ly$\alpha$ emitters at $z=8.96$ over an area of 1.4 deg$^2$. No Ly$\alpha$ emitter was found brighter than ${\rm L}\approx7.6\times10^{43}$\,erg\,s$^{-1}$.

\item The Ly$\alpha$ luminosity function constraints at $z\sim9$ have been improved for 10$^{42}<L<10^{45}$\,erg\,s$^{-1}$ emitters. The results rule out significant positive evolution of the Ly$\alpha$ LF beyond $z\sim6$; they are in line with recent semi-analytic \& phenomenological model predictions, rejecting some extreme models.

\item It has been shown that for narrow-band searches, T-dwarfs can mimic Ly$\alpha$ emitters at $7.7<z<8.0$, $9.1<z<9.5$ and $11.7 < z < 12.2$; they will not contaminate the future UltraVISTA narrow-band survey (and can even be identified via a narrow-band {\it deficit}), but they may contaminate narrow-band Ly$\alpha$ searches within the $z=7.7$ and $z=9.4$ atmospheric windows.

\end{itemize}

\section*{Acknowledgments}

The authors thank the reviewer, Jon Willis, for relevant suggestions. DS would like to thank the Funda{\c c}{\~ao para a Ci{\^e}ncia e Tecnologia (FCT) for the doctoral fellowship SFRH/BD/36628/2007. PNB acknowledges the Royal Society. JEG \& IRS thank the U.K. Science and Technology Facility Council (STFC) and KC acknowledges for a STFC Fellowship. The authors thank Cedric Lacey for providing the {\sc galform} model data and Eduard Thommes for sending updated phenomonological model results. The authors would also like to thank Andy Adamson, Chris Davies, Luca Rizzi and Tim, Thor and Jack for the support on the UKIRT telescope and acknowledge UKIRT service time.

\bibliographystyle{mn2e.bst}
\bibliography{bibliography.bib}


\bsp

\label{lastpage}

\end{document}